\documentstyle[12pt,moriond]{article}
\begin{document}
\heading{Statistical properties of local \& intermediate z galaxies}

\author{F. Hammer $^{1}$} {$^{1}$ DAEC, Observatoire de Paris Meudon, 92195 Meudon, France.} 

\begin{moriondabstract}
Galaxy evolution during the last 9 Gyr is discussed. It
can be traced back from well known present-day galaxies or directly
observed for galaxies at different look back times. This requires clear
and consistently matched selection criteria for galaxy samples.
There is a net decrease of rest-frame, UV luminosity density, at least
since z = 1. It is interpreted as an important decline of the star formation 
since the last $\sim$ 9 Gyr. A similar trend is found for the evolution of the IR
luminosity density which accounts for heavily extincted starbursts. Interestingly
the global star formation density, after including IR selected galaxies, is
twice the value of estimates based on the UV luminosity density, and this
holds from z=0 to z=1.\\  
Large disks are not contributing much to the observed decrease,
which is mainly related to significant changes with the epoch in the
distribution of galaxy morphologies. A significant fraction of the global star
formation occurs in luminous galaxies which are apparently small or in 
interacting galaxies detected in the deepest IR or radio surveys.

\end{moriondabstract}

\section{Introduction}

Studies of distant galaxies are limited by spatial resolution 
(HST/WFPC2, 0.1 arcsec or 1$h_{50}^{-1}$kpc at z$\ge$0.75) and by 
object faintness. The interesting features in the visible at rest are also
redshifted to the infrared window, where the OH atmospheric lines limit the
observations. For a galaxy redshifted at z=1.8-2.5, the visible window
virtually includes none of the strong emission lines. There are 
 two important reasons for further studying z $\le$ 1 galaxies:\\
 -a large fraction of the present day stellar mass has been formed since the last
 9 Gyr (z$\sim$1);\\
 -they are bright enough for optical/near IR detailed studies aimed to 
 understand their morphology (luminosity profiles), their chemistry \&
 their dynamics.\\
The aim of this
 review is to summarize the most recent progresses in understanding galaxy
 evolution since z=1. Observational selection effects have to be carefully
 examined as they can easily mimic evolutionary trends.
Distant galaxies are naturally
 selected by their observed fluxes, which lead to various selection biases
 in spectroscopic samples:\\
-imagery used for selecting sources should be at least 1-2 magnitude deeper
than the spectroscopic limit in order to securely include galaxies with low surface
brightness.\\
-high redshift galaxy samples, if selected in the visible -i.e. rest-frame UV-,
are obviously biased against both the old stars and the obscured massive stars; this
 has motivated some selections in the I-band at 0.835$\mu$m (up to z=1) or K band 
 at 2.2$\mu$m (up to z=4); also IR or radio selections is a prerequisite to a
 proper evaluation of the energy output reprocessed by dust.\\
-deep pencil beams based on too small areas (i.e. few 
square arcmin) are not ideal to sample galaxy evolution; they 
include galaxies within a large redshift range and redshift dependent effects
could become rather complex; in the lowest redshift bin they sample a too small volume,
including only galaxies with much lower luminosities than at a higher redshift.\\
 The latter point emphasizes the need for fair comparison samples at very low-z,
 which represent the non-evolution reference. Selection at low z is far from
 being trivial, and can be also affected by large scale structures.\\
  General properties of nearby 
 galaxies will be presented, including local luminosity \& 
 star formation density, as well as
 the star formation processes in disk and in 
 circumstellar nucleus regions. Evolution of 
quantities at various look-back times will be examined, including
 star formation density derived from UV, mid-IR \& radio light density 
measurements. This leads to a general discussion on how extinctions can
limit our view of galaxy evolution. Finally,
 star formation can be differentially traced for galaxies of various
morphological types.     
Galaxies show a wide variety in size, mass, morphology, overall
energy distribution, nuclear activity etc... Understanding the relative
contribution of a galaxy class to the star formation density requires
large samples (100 objects provide a Poisson uncertainty of 10\%) selected with
 a reliable criterion within the available redshift range.

\section{General properties in local galaxies}
\subsection{Luminosity densities and star formation density}

 The difficulty in establishing
global properties of local galaxies is illustrated by the long debate about the 
exact shape luminosity function (see Efstathiou et al, 1988). There is now a
general agreement that the blue luminosity function is rather steep ($\alpha$=
-1.2) at its faint end (Zucca et al, 1997). Converted into blue luminosity
density this provides: $(\phi L)_{B})$=(3.9 $\pm$0.5) $10^{7}$ $h_{50}$ $L_{\odot}$ $Mpc^{-3}$
(Loveday et al, 1992; Marzke et al, 1994).
Integration of the K band luminosity function gives: $(\phi L)_{K})$=
(4 $\pm$ 1) $10^{8}$ $h_{50}$ $L_{\odot}$ $Mpc^{-3}$ (Gardner et al, 1997). This
leads to a stellar mass density of (4 $\pm$ 2) $10^{8}$ $h_{50}^{2}$ $M_{\odot}$ $Mpc^{-3}$,
assuming 0.6$h_{50}<$ $M/L_{K}$ $<$ 1.8$h_{50}$ and a Salpeter IMF. The average 
color of the local stellar population taken as a whole is then 
$(B - K)_{AB}$ = 2.5, a value typical for a Sab galaxy.\\

Gallego et al (1995) have selected 
z$\le$0.045 galaxies from $H\alpha$ emission in Schmidt objective-prism plates. 
According to their results, the $H\alpha$ luminosity density of local 
Universe is $10^{39.1\pm0.04}$$h_{50}$$ erg s^{-1}Mpc^{-3}$. This leads to a
SFR density of 0.008 $\pm$ 0.0006 $h_{50}$ $M_{\odot}yr^{-1}Mpc^{-
3}$. Using the same technics, Gronwall et al (1998, KISS project) suggest 
twice this value. 
Based on UV selected galaxies, Treyer et al (1998) have
derived $9.3_{-0.45}^{+0.76}$ $10^{25}$$h_{50}$ erg $s^{-1}Hz^{-1}Mpc^{-3}$ for 
the luminosity density at 2000\AA~of z$\sim$ 0.15 galaxies. This corresponds to 
a SFR density of $0.012_{-0.003}^{+0.005}$ $h_{50}$ $M_{\odot}yr^{-1}Mpc^{-3}$,
i.e. 50\% higher than the Gallego et al value. All the above estimates are 
assuming a Salpeter IMF (0.1-125$M_{\odot}$), and without correction for
 dust extinction. One would expect the $H\alpha$ estimated SFR to be less
 affected by biases against dusty objects, hence higher than UV estimates.
  One can suspect overestimation of SFRs for
a substantial fraction of Treyer et al galaxies, which are extremely blue
and are probably young starbursts with low metallicities. \\
\\
Tresse and Maddox (1998) have
estimated the $H\alpha$ luminosity density of z$\sim$0.2 galaxies, 
assuming an average 0.45 mag extinction correction at 6562A. This is still
an uncertain estimate. Applying this to the Gronwall et al (1998) value
 would provide a local density of star
formation of  0.024 $h_{50}$ $M_{\odot}yr^{-1}Mpc^{-3}$. It is probably 
irrelevant to apply the same correction to FOCA UV-selected galaxies, since
they are likely to be less dusty than $H\alpha$ galaxies.\\
 
On the other hand, IRAS galaxies contribute only marginally 
 to UV  
light density, while they individually produce large amounts of bolometric
 luminosity and/or star formation. Estimates of local SFR density based on 
 IRAS measurements (Saunders et al, 1990), provide
a SFR density of $\sim$ 0.012 $h_{50}$ $M_{\odot}yr^{-1}Mpc^{-3}$, i.e. 
close to the FOCA value. So the exact value of the local SFR density 
 is probably slightly below twice the IRAS
 value (0.024 $h_{50}$ $M_{\odot}yr^{-1}Mpc^{-3}$). Locally there is an apparent
 equipartition of the energy balance between UV and IR emissions from hot stars.

\subsection{Star formation in disks}

Star formation currently occurs in galactic disks, and is generally 
estimated from $H\alpha$ and UV flux measurements. It varies considerably
from one galaxy to another, from virtually 0 in gas poor S0 to 1 $M_{\odot}yr^{-
1}$ in our Galaxy, and to $\sim$ 10 $M_{\odot}yr^{-1}$ in
gas rich disks. It is more convenient to compare the star formation rate per
unit of red luminosity, because galaxies show a large range in mass/luminosity.
This parameter -SFR/$<$SFR$>$, where $<$SFR$>$ is the past averaged SFR-,
provides a good representation of the relative strength of star formation
in a galaxy. It is generally estimated from the $H\alpha$ equivalent width
(Kennicutt et al, 1994). The strength of star formation increases on average, 
from early type (Sa: $\sim$ 0.1) to late type (Sc-Sd: $\sim$
1). The dispersion within a morphological class is however very large
(factor $\sim$ 10), reflecting the large spread of galaxy properties 
and histories.\\
Relating the star formation surface density to the gas content (Schmidt law),
Buat (1992) and Kennicutt (1998) have derived low time scales (2-5 Gyr) for the
gas consumptions. This is consistent with findings based on our own Galaxy
(see Prantzos and Aubert, 1995).

\subsection{Star formation in circumnuclear regions}

Star formation also occurs in compact regions, generally in galaxy nuclei.
Nuclear HII regions are found in 42\% of bright spirals, with the fraction
increasing from S0 to Sc-Im galaxies (Ho et al, 1997). They show modest
SFRs, averaging from 0.1-0.2 $M_{\odot}yr^{-1}$. However SFRs
can reach values in excess of 100 $M_{\odot}yr^{-1}$ in the much less 
numerous population of luminous IRAS galaxies. It might be particularly
 difficult to disentangle star formation
from AGN emission in the ultra-luminous IRAS galaxies (ULIRG). Lutz et al
(1998) found 50\% of AGN in ULIRGs with $L_{IR}>$ 2 $10^{12}L_{\odot}$.\\
\\
SFRs in luminous IRAS galaxies are generally estimated from their FIR 
luminosity,
assuming that most of the ionizing photons are re-processed by dust. 
In these dense regions, most of the gas is molecular (see Sanders and 
Mirabel, 1996). Kennicutt (1998) found that Schmidt law is followed more
tightly by compact starburst galaxies
than by disks, though sharing the same slope. The global star formation 
efficiencies are much higher than in 
normal disks, with a characteristic time scale for gas consumption of
few tenths of Gyr.

\section{Evolution of star formation from deep surveys of galaxies}
\subsection{Evolution of global quantities at optical wavelengths}

The most exhaustive study of the galaxy evolution up to z=1 is provided
by the Canada France Redshift Survey (CFRS, Lilly et al, 1995) which
includes 600 galaxies with 0.1$\le$ z $\le$1. From its
selection criterion ($I_{AB}\le$ 22.5), all the $M_{B}(AB) \le$ -20 galaxies are
 included in the sample, and this up to z=0.9-1. Hammer et al (1997) found
that the fraction of star forming galaxies increases with the redshift: more 
than 50\% at z$>$0.5 have $W_{0}(OII)>$15A, which should be 
compared to 13\% locally (see Vettolani et al, 1998). This observed trend for
CFRS luminous galaxies ($M_{B}\le$-20) is followed by a shift in color: the average
rest-frame $(U - V)_{AB}$ color varies from 1.6 (Sab color) at z$\sim$0,
to 1.3 (Sbc color) at z=0.5 and 0.7 (Sdm-Irr color) at z=1.\\
\\
The increase in star formation with redshift has been quantitatively 
estimated by Lilly et al (1996) from the rest frame 2800\AA~luminosity, whose 
comoving density evolves as rapidly as $(1 + z)^{3.9\pm0.75}$. This value
is provided after assuming a constant slope of the luminosity function
in the $[0,1]$ redshift range. A
similar, even stronger trend is also observed for the [OII]3727 comoving 
luminosity density, which follows $(1 + z)^{6.5\pm2.5}$ (Hammer et al, 1997). 
Differences between exponents
could be due to errors in measuring [OII] fluxes in faint and distant
galaxies, to extrapolation of 2800\AA~luminosities at low redshift, and possibly
to changes with the redshift in the average metal abundance in galaxies.
These evolutionary changes have been interpreted as due to a large decrease of the star
formation by a factor 10 from z=1 to z=0 (Madau et al, 1996). According to 
them and to Steidel et al (1996), this suggests 
a peak at z=1-2 in the comoving UV luminosity density.\\

\begin{figure}
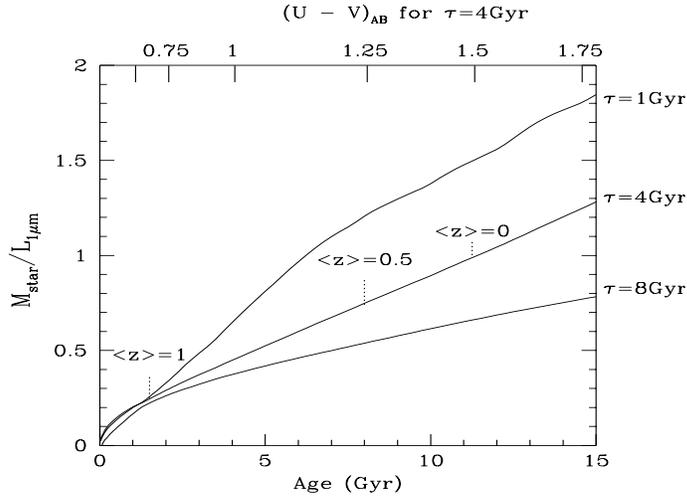

  \begin{center}
    \leavevmode
  \end{center}
  \caption{ {$M_{star}/L_{1\mu}$ versus galaxy age from Bruzual and Charlot
(1999) stellar tracks with SFR $\sim$ $exp(-t/\tau)$. An average present-day
 age of 11.25 Gyr has been assumed for luminous galaxies observed in the CFRS
sample. At larger redshifts galaxies are younger on average, they are bluer
($(U-V)_{AB}$ color is displayed on the top) and produce more $1\mu$m light
per unit mass.}}
\label{figure1}
\end{figure}
   
Lilly et al (1996) also found a redshift increase of the comoving 1$\mu$m
 light density, following $(1 + z)^{2.1\pm0.5}$. This could
mainly explained by the expected shift in age and color of the
 bright galaxies (see Figure 
1). At z=1, galaxies were on average younger by 8-9 Gyr, their 
$(U - V)_{AB}$ colors were bluer by $\sim$ 1 mag, and their 
$L_{1\mu}/M_{star}$ 
ratio were larger by a factor of $\sim$ 4 at z=1 than today.

\subsection{Evolution of spectral properties}

Hammer et al (1997) found that 40\% of the CFRS emission line galaxies show 
spectra with features typical of an important A star population (Balmer 
continuum or $W(H\delta)$). There is no trend with the redshift, 
and this is consistent with $W(H\delta)$=5\AA~ found by Kennicutt (1992) in
a sample of local galaxies. Population of A star is generally thought to be a 
reminiscence of a previous burst occurring few tenths of Gyr ago. At z$\ge$0.5,
many galaxies simultaneously present current star formation and A star population.
This suggests that most galaxies have experienced long periods of
star formations ($>$ 1Gyr) or, alternatively, numerous consecutive
small bursts during long periods. In addition, only few galaxies ($\sim$ 5\%) 
present an HII flat spectra.\\
\\
Redshifted galaxies show a wide range of ionization properties (Hammer
et al 1997). Beyond z=0.7, 30\% of the CFRS galaxies have 
continuum properties (namely their relation between 4000A break intensity 
and UV continuum slope), not reproducible by population synthesis models with 
solar abundance. Their continua are similar to those of low abundance 
Magellanic 
star clusters, suggesting low metallicities for a significant fraction of
star forming galaxies at z$\ge$0.7.

\subsection{Calibration of star formation for redshifted galaxies}

The choice of star formation tracers in the visible, is limited when looking
at a population of redshifted galaxies. $H\alpha$ line is observable up to
 z=0.5, and [OII]3727 up to z=1.5. On the other hand, visible observations
of redshifted galaxies are measuring rest-frame UV fluxes of galaxies which 
could be a direct measure of the star formation. However, both [OII]3727 and
2800\AA~ luminosities appear to be poor tracers of the star formation, because 
they are not well correlated with $H\alpha$ luminosity (Hammer and Flores, 1998). 
Since [OII]3727 and 2800\AA~ luminosities correlate well together,
this suggests that extinction is the major source of uncertainties when 
these luminosities are used for tracing star formation. This is illustrated in 
Figure 2, which shows the relation between $W(H\beta)$ and $W(H\alpha)$.
 This relation is much more dispersed than
the one of Kennicutt (1992) for local galaxies. This is not
 necessarily related to an evolutionary effect,
and could be simply a selection effect. For example, Jansen et al
(1999) provided spectrophotometric measurements of $\sim$ 200
local galaxies representative of the local luminosity function and found a
 relation between $W(OII)$ and $W(H\alpha)$ with a dispersion similar to
 our result at higher redshift.\\

\begin{figure}
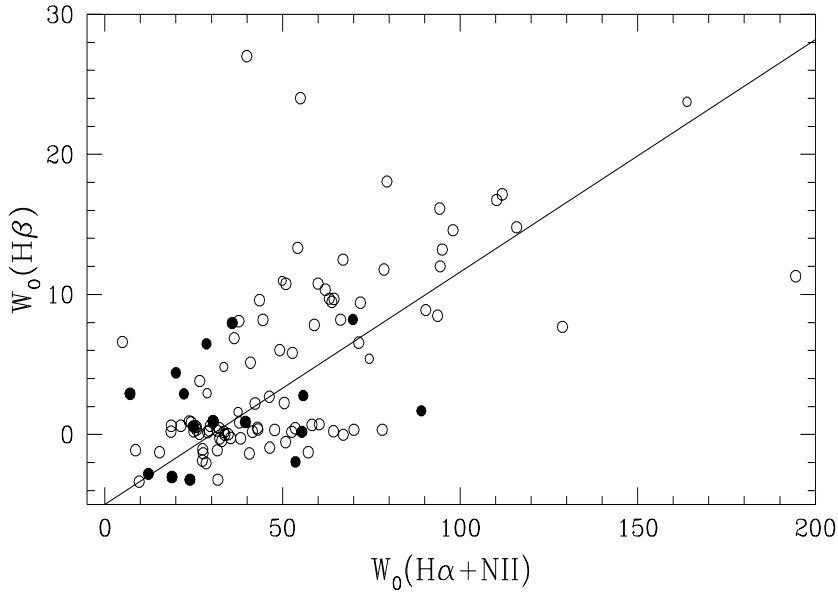

  \begin{center}
    \leavevmode
  \end{center}
  \caption{ {$W(H\beta)$ versus $W(H\alpha+NII)$ for 100 CFRS galaxies with 0.15$\le$ z $\le$0.5. 
  Solid line shows the tight correlation found by Kennicutt (1992) in
his local sample. Full and open dots represent $M_{B}\le$ -20 and $M_{B}>$ -20,
repectively.}}
\label{figure1}
\end{figure}
\normalsize

Most of the galaxies which lie above the Kennicutt (1992) fiducial relation have
colors bluer than a Sbc ($(U - V)_{AB}\le$1.3). A crude interpretation of Figure 2
 would be that these galaxies suffer from extinctions similar
to those of local irregular galaxies (i.e. $A_{V}\sim$0.6, Gallagher et al 1989)
, or two times less than as derived for Kennicutt galaxies ($A_{V}$=1.2). 
This value is
consistent with findings at z$\sim$ 0.2 by Tresse and Maddox (1998), at 
z$\sim$ 1 and at z$\sim$ 3 
(Pettini et al, 1998). One should however notice (see Figure 2) the
presence of galaxies with non negligible $H\alpha$ lines and with virtually
no $H\beta$, [OIII]5007 and [OII]3727 lines. These could be very extincted galaxies 
which might affect 
estimates of SFRs based on $H\alpha$ lines (Gruel et al,
 1999, in preparation).
 
\subsection{Star formation evolution seen in infrared and at radio wavelengths}

Observations in the visible of redshifted galaxies are likely
biased against star forming, dust enshrouded galaxies. Recent ISOCAM deep
surveys show that 15$\mu$m counts below 1mJy present a significant excess
in galaxy numbers relatively to no-evolution models (Elbaz et al, 1999).
Strong starbursts forming 50 $M_{\odot}yr^{-1}$ at z $\sim$ 1 can
be easily detected by ISOCAM deep exposures and also by VLA deep surveys
(Flores et al, 1999). \\

From a CFRS follow-up study with ISOCAM, Flores et al (1999) have provided a 
first estimate of the fraction of
 star formation density which is hidden by dust. 
Classification of $\sim$ 30 sources with 0.2$\le$ z $\le$1 has been done from
the spectral energy distribution based on rest-frame UV, visible, near and mid-IR
 and radio photometric points, from line diagnostic diagrams and from radio
slopes and sizes. It results that 60\% of the $S_{15\mu}\ge$ 250$\mu$Jy
sources are starbursts, 25\% are Seyfert2 or Liners, and 15\% are 
broad-line emission objects. ISOCAM galaxies at 0.5 $\le$ z $\le$ 1 have
IR luminosities lower than 2 $10^{12}$ $L_{\odot}$, and the fraction of 
AGN is significantly higher than what Lutz et al (1998) found locally 
($\sim$ 15\%). At least a part of the discrepancy could be a selection effect,
 because Lutz et al (1998) selected
galaxies from their fluxes at 60$\mu$m, a wavelength which is much more
sensitive to starburst galaxies than the Flores et al 15$\mu$m/(1+z) limit.\\ 

\begin{figure}
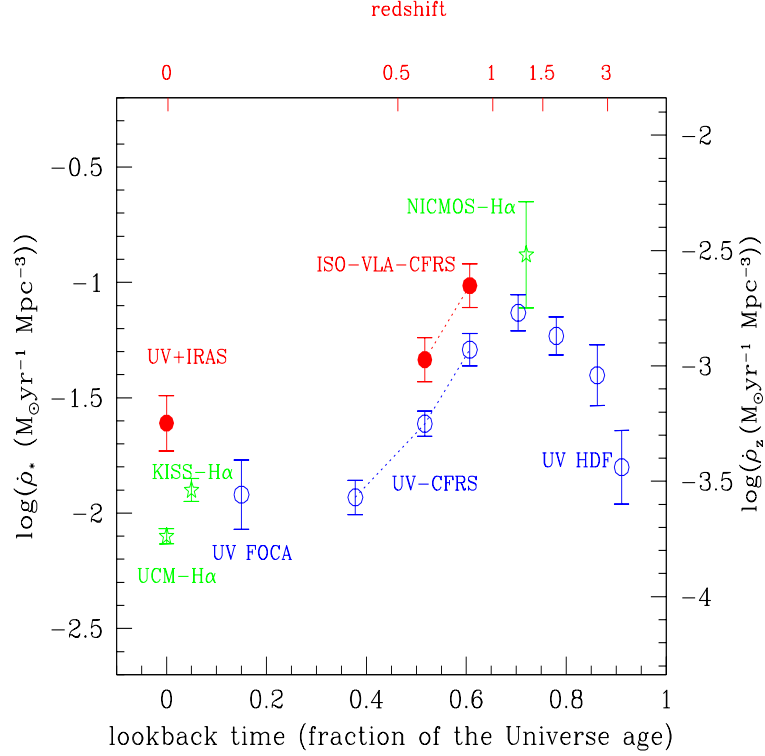

  \begin{center}
    \leavevmode
  \end{center}
  \caption{ {Metal production and star formation history. SFR estimates are 
  assuming a Salpeter IMF from
0.1 $M_{\odot}$ to 100 $M_{\odot}$. Flores et al
  points (filled circles, labeled ISO-VLA-CFRS) are 1.9 times higher 
in SFR density or in metal production than those 
(open circles) previously derived from 
the UV flux density at 2800\AA. The same situation is found locally with
an equal contribution from IRAS galaxies and UV selected galaxies (FOCA, 
Treyer et al, 1998) to the star formation density. At higher z, estimates
are derived from UV rest frame 
wavelengths and come from Connolly et al (1997) and Madau et al (1996) (HDF) 
(open squares). No UV estimate has been corrected for extinction. 
 Other estimates (stars) from $H\alpha$ are given for comparison and are from 
 Gallego et al (1995, UCM) and Gronwall et al (1998, KISS), as well as at
 higher z from few galaxies at z$\sim$ 1.25 by Yan (1999).}}
\label{figure1}
\end{figure}

Flores et al (1999) conclude that 4\% of
the field galaxies are strong and heavily extincted starbursts with
SFR from 40 to 200 $M_{\odot}yr^{-1}$, and produce a third of the global
star formation density at z $\sim$ 1. To provide the
corresponding star formation density, one could
assume no evolution for the low-end slope of the IR luminosity function as it was done
 by Lilly et al (1996) for UV estimates. It results in a global star formation 
density 1.9$\pm$0.7 times higher than 
that derived from UV measurements (Figure 3). Error bars are still large and
are related to small statistics as well as to the ambiguity about the source 
of IR emission in some luminous galaxies (Seyfert2). The amplitude of the
SFR density evolution at IR wavelengths is apparently similar to that estimated
at UV wavelengths. Most of the ISOCAM sources at 0.5$\le$z$\le$ 1,
appear to be strong mergers, or at least they show signs of interactions.

\section{Morphology evolution}
Morphological studies of redshifted galaxies are complicated by 
redshift dependent effects. For example, the
commonly used I broad band filter ($I_{814W}$ HST/WFPC2) is sampling the 
rest-frame B band at z=0.9, a color which is more sensitive to star
forming regions than in the redder ones. This effect has been extensively 
studied by Brinchman et al (1998). They quote that 24\% of the spirals at 
z=0.9 would be 
mis-classified as irregular because of that redshift effect. This is likely
to become predominant at higher redshift and could severely affect the conclusions on
 morphological evolution studies. Another expected bias in selecting
galaxies in optical could be related to a deficiency in edge-on galaxies, which
are more affected by extinctions. Several distant galaxies are so compact
that they are merely resolved with HST/WFPC2. This probably provides the most
severe limitation to the morphological studies of distant galaxies.\\

Brinchman et al (1998) have presented the HST imagery of $\sim$ 340 galaxies up to
z=1, using the CFRS selection criterion. Figure 4 shows the 
{\it morphology-color} evolution of the field galaxies from z=0.25 to z=1. 
The color evolution discussed in section 3.1 is
well adjusted to a general shift towards later 
types at higher redshifts. Brinchman et al (1998) quoted that 9\% of galaxies
at 0.2$<$ z $<$0.5 are irregular, a fraction which reaches
32\% at 0.75$<$ z $<$1. Indeed, extended star forming galaxies ($W_{0}(OII)\ge$ 
15\AA~) generally show irregularities, companions or circumstellar regions not
always located at the galaxy centers.\\

\begin{figure}
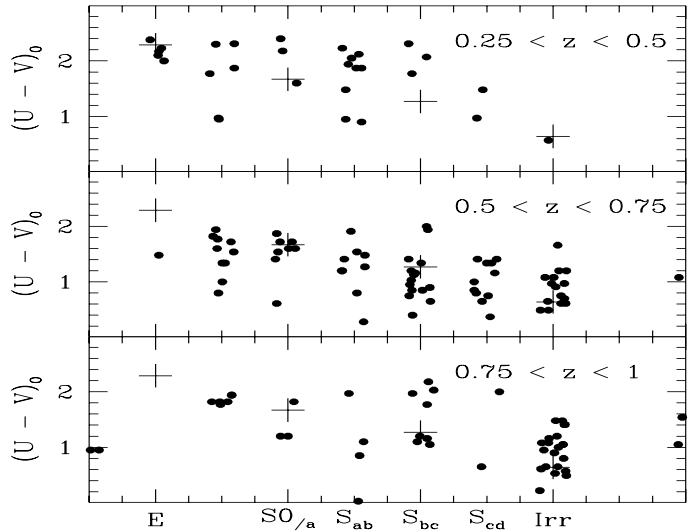

  \begin{center}
    \leavevmode
  \end{center}
  \caption{ {Rest frame $(U - V)_{AB}$ color versus HST morphological
class for $M_{B}\le$ -20 CFRS, in three redshift bins. Brinchman et al
classification is supported by the comparison with local values (large
crosses; Coleman et al, 1980). Most galaxies have an earlier type than Sbc in the 
lower redshift bin, while the reverse is found at high redshift.}}
\label{figure1}
\end{figure}

The density of large disks
 with $r_{disk}\ge$ 3.2$h_{50}^{-1}$kpc is found to be the same at z=0.75 than
locally (Lilly et al, 1998). Only a density decrease by less than 30\% 
at z=1 is consistent with the data. Lilly et al (1998) also find that 
star formation in large disks present only a modest increase with the redshift.
 From long-slit spectroscopy studies, Vogt et al (1997,
see also Koo et al, in this volume) show an almost unevolved Tully Fischer
relation for disks at z $\sim$ 1. Redshift changes in large disks appear not to
be the main contributors to star formation evolution as detected in
that redshift range. At large distance most of the disks are of a late type.\\

The most rapidly evolving population of galaxies is made of
small and compact galaxies (Lilly et al, 1998, see also Guzman, in this volume) . This 
confirms the Guzman et al (1997) claim, based on a HDF sample in the
same redshift range. Their UV luminosity density was 10 times higher 
at z=0.875 than at
z=0.375, and they correspond to $\sim$ 30\% of the rest-frame UV luminosity
density in the higher redshift bin (Hammer and Flores, 1998b). These objects 
are somewhat enigmatic: their sizes -$r_{disk}\le$ 2.5$h_{50}^{-1}$kpc- and
their velocity widths -35 to 150 km/s (Phillips et al, 1997)- are apparently similar 
to those of local dwarves, while they are up to 100 times more luminous 
than a $M_{B}$=-17.5 dwarf.

\section{Conclusion}

There is a general decrease of star formation in galaxies since at least the
last 9 Gyr (z$\sim$ 1). This is supported by:
\begin{itemize}
\item a general decrease of the rest-frame UV \& IR luminosity density, since z=1
\item average properties of z$\sim$ 1 luminous galaxies are broadly
 consistent with those of blue, starbursting and irregular galaxies.
\end{itemize}
A large fraction of galaxies in the past show an important population of
A stars, and very few galaxies show a HII-like spectrum. So star formation
in most field galaxies appears to be a continuous process during long
periods of times, or alternatively to be the result of numerous and successive
bursts.\\
 Large disks present a modest evolution since the last 9 Gyr. 
Star formation in these systems is sustained over time scale $\le$ 5 Gyr. These results
are solid because they are independently obtained from analysis of nearby
and distant galaxies. Most of the reported star formation evolution found
in the UV seems to be related to a population of star forming, compact galaxies. 
At a high redshift they have sizes and velocities apparently comparable
to those of local dwarves, while they are overluminous by factors reaching 100.\\
Uncertainties on dust extinctions appear to be a major problem in 
estimating the evolution of star formation. Accounting for the population
of obscured, strong starbursts detected
at IR and radio wavelengths would increase the SFR density by a factor
1.9$\pm$0.7, when compared to estimation based on UV luminosity
density. From the ratio of IR to UV luminosity density, one can derive
a moderate global extinction at 0$\le$ z $\le$ 1, corresponding 
to $A_{V}$=$0.45^{+0.3}_{-0.2}$.\\
 From z=0 to z=1, the global energy output 
from hot stars seen directly at UV wavelengths is similar to that reprocessed
by dust. This result is consistent with the bolometric measurements from UV to
sub-millimeter wavelengths which integrate all the energy emitted by
extragalactic sources (see Pozetti et al, 1998). Integrating global star
formation from z=1 to z=0 provides a value from 50\% to 100\% of present
day stellar mass. This might prevent from a scenario in which 
 star formation density is still increasing beyond z=1-2. However, a large fraction 
 of the stellar mass and metal lie in metal rich bulges, and the next important 
 challenge is to know when most of these objects have been formed.

\vfill
\end{document}